\documentclass{aa}
\usepackage{graphics}
\def\a{{$\alpha$}}

\def\gsnr{{G~114.3+0.3}}
\newcommand{\h}{$^{\rm h}$}
\newcommand{\m}{$^{\rm m}$}
\newcommand{\s}{$^{\rm s}$}
\newcommand{\dd}{$\delta$}
\newcommand{\ha}{\rm H$\alpha$}
\newcommand{\hbeta}{\rm H$\beta$}
\newcommand{\HII}{\ion{H}{ii}}
\newcommand{\hnii}{{\rm H}$\alpha+[$\ion{N}{ii}$]$}
\newcommand{\nii}{$[$\ion{N}{ii}$]$}
\newcommand{\sii}{$[$\ion{S}{ii}$]$}
\newcommand{\oii}{$[$\ion{O}{ii}$]$}
\newcommand{\oiii}{$[$\ion{O}{iii}$]$}
\newcommand{\snr}{\rm supernova remnant}

\newcommand{\et}{et al.}
\newcommand{\flux}{$10^{-17}$ erg s$^{-1}$ cm$^{-2}$ arcsec$^{-2}$}
\newcommand{\dens}{\rm cm$^{-3}$}
\newcommand{\vel}{\rm km s$^{-1}$}
\begin{document}

%   \thesaurus{08     % A&A Section 8: Diffuse matter in space
%              (09.07.01;  % ISM : general,
%		09.19.2;
%		09.09.1)} % Superona remnants
%
\title{Imaging and spectroscopy of the faint remnant G 114.3$+$0.3}

\author{F. Mavromatakis\inst{1}
\and  P. Boumis\inst{1}
\and E. V. Paleologou\inst{2}
}
\offprints{F. Mavromatakis,\email{fotis@physics.uoc.gr}}
\authorrunning{F. Mavromatakis et al.}
\titlerunning{Optical observations of \gsnr}
\institute{
University of Crete, Physics Department, P.O. Box 2208, 710 03 Heraklion, Crete, Greece 
\and Foundation for Research and Technology-Hellas, P.O. Box 1527, 711 10 Heraklion, 
Crete, Greece}
%\date{Received ...... / Accepted .......}
\date{Accepted }

\abstract{
We present the first calibrated CCD images of the faint supernova remnant 
\gsnr\ in the emission lines of \oii, \oiii, \hnii\ and \sii. 
The deep low ionization CCD images 
reveal diffuse emission in the south and central areas of the remnant. 
These are correlated with areas of intense radio emission, while 
estimates of the \sii/\ha\ ratio suggest that the detected emission 
originates from shock heated gas. 
In the medium ionization image of \oiii\ we discovered a thin filament in the 
south matching very well the outer radio contours. This filament is not
continuous over its total extent but shows variations in the intensity, 
mainly in the south--west suggesting inhomogeneous interstellar clouds. 
Deep long--slit spectra were also taken along the \oiii\ filament clearly 
identifying the observed emission as emission from shock heated gas. 
The \ha\ emission is a few times \flux, while the variations seen in the 
\oiii\ flux suggest shock velocities into the interstellar clouds 
around or below 100 \vel. The sulfur line ratio approaches the low density
limit implying electron densities less than $\sim$500 \dens.
}
\maketitle
\section{Introduction}
The vast majority of supernova remnants have been discovered by their 
non--thermal synchrotron radio emission. 
Since the optical wavelengths suffer significantly more attenuation than the 
radio, the detection of a \snr\ in the optical band is a difficult task. 
The \snr\ \object{G 114.3+0.3} was initially detected in the 21 cm continuum survey of 
Kallas \& Reich (\cite{kal80}) and subsequently studied in more detail 
by Reich \& Braunsfurth (\cite{rei81}). The remnant shows up in the radio as 
an ellipsoidal shell occupying an angular extent of $\sim$60\arcmin\ $\times$ 
78\arcmin. The radio emission is stronger in the south--east and 
south--west compared to other areas of the remnant. The surface brightness 
is low, while at the same time the source displays a high degree of 
polarization. More interest was  raised for this system following the 
proposal of F\"urst \et\ (\cite{fur93}) and Kulkarni \et\ (\cite{kul93}) that the 
pulsar PSR 2334$+$61 is associated with \object{SNR 114.3+00.3}. 
This pulsar rotates at a period of 495 ms and its spindown rate suggests an 
age of $\sim$ 41000 yr. The distance estimates to this pulsar--remnant 
association lie in the range of 2--3 kpc. ROSAT PSPC pointed observations 
showed that the pulsar is a weak X-ray source, while no emission was 
detected from the remnant itself (Becker \et\ \cite{bec96}).
\par
Fesen \et\ (\cite{fes97}) reported the detection of faint filamentary 
structures in \ha\ in the west, south--west areas of the remnant. 
However, no images were shown due to the faintness of the filaments.
Even though the spatial correlation of the optical emission with the 
radio contours favors their association, flux calibrated images in \ha\ and 
\sii\ or long--slit spectra are needed to establish the nature of the optical
flux. In an effort to deepen our knowledge of the properties of the optically 
detected remnants, especially the faintest ones, we performed deep CCD 
imaging and spectral observations of \gsnr. 
Information about the observations and the data 
reduction is given in Sect. 2. In Sect. 3 and 4 we present the results of 
our imaging observations and the results from
the long--slit spectra taken at specific locations of interest. 
Finally, in Sect. 5 we discuss the physical properties of the remnant.
\section{Observations}
\subsection{Optical images}
The observations presented in this paper were performed with the 0.3 m 
Schmidt Cassegrain telescope at Skinakas Observatory, Crete, Greece. 
The field of \gsnr\ was observed in July 12 and 14, 1999. 
The observations were performed with a 1024 $\times$ 1024 Site CCD which 
had a pixel size of 25 $\mu$m resulting in a 89\arcmin\ $\times$ 89\arcmin\ 
field of view and an image scale of 5\arcsec\ per pixel. 
The number of frames taken is given in Table~\ref{obs} along with the total 
exposure times in the specific filters used during the observations. 
The filter characteristics can be found in Mavromatakis \et\ (\cite{mav00}).
The final images in each filter are the average 
of the individual frames where appropriate, while during the astrometry 
process the HST Guide star catalogue was used. All coordinates quoted in this 
work refer to epoch 2000.
\par
We employed standard IRAF and MIDAS routines for the reduction of the data. 
Individual frames were bias subtracted and flat-field corrected using 
well exposed twilight flat-fields. The spectrophotometric standard stars 
HR5501, HR7596, HR7950 and HR8634  were used for absolute flux calibration.
\subsection{Optical spectra}
Long--slit spectra were obtained on July 16, 21, and 22, 2001  using the 
1.3 m Ritchey--Cretien telescope at Skinakas Observatory. 
The data were taken with a 1300 line mm$^{-1}$ grating 
and a 800 $\times$ 2000 Site CCD covering the range of 4750 \AA\ -- 6815 \AA.
The slit had a width of 7\farcs7 and, in all cases, was oriented
in the south-north direction. The coordinates of the slit centers, the 
number of available spectra from each 
location and the exposure time of each spectrum are given in 
Table~\ref{spectra}. 
The spectrophotometric standard stars HR5501, HR7596, HR9087, HR718 and 
HR7950 were observed in order to calibrate the spectra of \gsnr.  
\begin{table}
      \caption[]{Log of the exposure times}
         \label{obs}
\begin{flushleft}
\begin{tabular}{lllllll}
            \noalign{\smallskip}
\hline
  \hnii\	& \sii\		& \oiii\ & \oii\    \cr
\hline
3600$^{\rm a}$(2)$^{\rm b}$	&3600 (2)&6600(3)&1800(1)	 \cr
 \hline
\end{tabular}
\end{flushleft}
${\rm ^a}$ Total exposure time in sec \\\
${\rm ^b}$ Number of individual frames \\\
   \end{table}
  \begin{table}
      \caption[]{Spectral log}
         \label{spectra}
\begin{flushleft}
\begin{tabular}{lllllll}
            \noalign{\smallskip}
\hline
	Slit centers &  &No of spectra & \cr
\hline
	 $\alpha$ & $\delta$ & (Exp. times)  &  \cr
\hline 
 23\h39\m46\s & 61\degr21\arcmin43\arcsec 	& 2$^{\rm a}$ (3600)$^{\rm b}$ \cr
\hline
 23\h39\m17\s & 61\degr19\arcmin06\arcsec 	& 2$^{\rm a}$ (3600)$^{\rm b}$ \cr
\hline
 23\h36\m37\s & 61\degr17\arcmin41\arcsec 	& 2$^{\rm a}$ (3600)$^{\rm b}$ \cr
 \hline
\end{tabular}
\end{flushleft}
${\rm ^a}$ Number of spectra obtained \\\
${\rm ^b}$ Exposure time of individual spectra in sec\\\
   \end{table}
  \begin{table}
      \caption[]{Typically measured fluxes}
         \label{fluxes}
\begin{flushleft}
\begin{tabular}{lllll}
            \hline
            \noalign{\smallskip}
     & SE$^{\rm a}$  & SW$^{\rm a}$ & Center$^{\rm a}$ & LBN 565$^{\rm b}$\cr	
\hline
\hnii\ 	& 11	& 10  	& 14	&   90    \cr	
\hline
\sii\ 	& 2.4	& 3.5	& 2.6  &    9	\cr	
\hline
\oiii\ 	& 1.0	& 0.6	& --   &  -- \cr	
\hline
\oii\ 	& 2	& 1.6	& --  &  5   \cr
\hline
\end{tabular}
\end{flushleft}
${\rm }$ Fluxes in units of \flux \\\
$^{\rm a}$Median values over a 112\arcsec $\times$ 67\arcsec\ box \\\
$^{\rm b}$Median values over a 394\arcsec $\times$ 394\arcsec\ box \\\
 \end{table}
\section{Imaging of \gsnr}
  \begin {figure}
   \resizebox{\hsize}{!}{\includegraphics{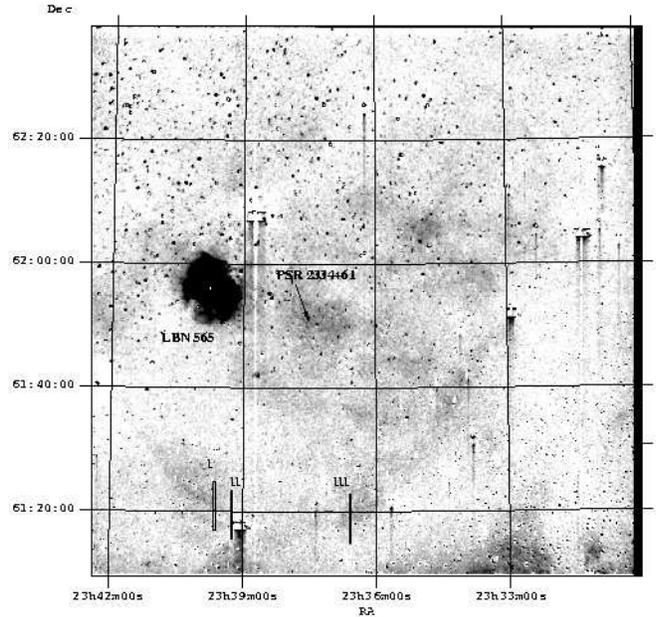}}
    \caption{ The field of \gsnr\ in the \hnii\ filter. 
     The image has been smoothed to suppress the residuals 
     from the imperfect continuum subtraction.
     Shadings run linearly from 0 to 20$\times$ \flux. The long rectangles
     show the positions observed through long--slit spectroscopy.
     The line segments seen near over-exposed stars in this 
     figure and the next figures are due to the blooming effect.
      } 
     \label{fig01}
  \end{figure}
%--------------------------------------------------------
  \begin {figure}
   \resizebox{\hsize}{!}{\includegraphics{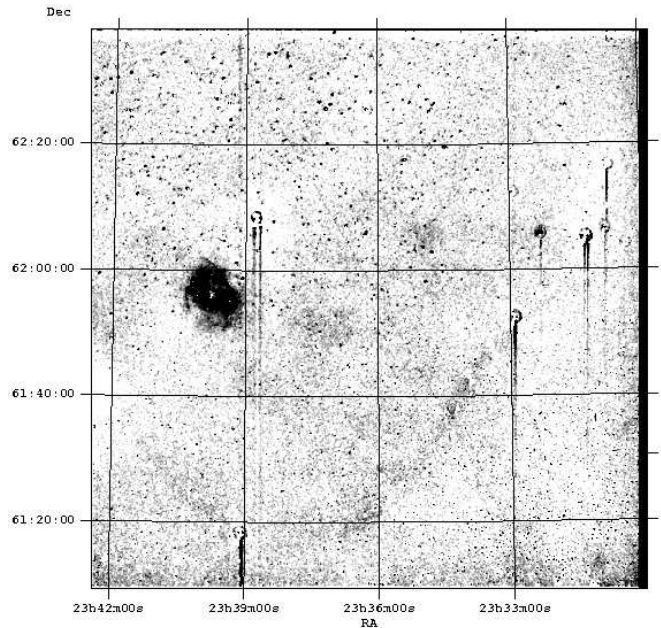}}
    \caption{ The \sii\ image of the area around \gsnr. 
     The image has been smoothed to suppress the residuals 
     from the imperfect continuum subtraction. Note the very strong 
     \HII\ region LBN 565 in the central east area of the remnant.
     Shadings run linearly from 0 to 10$\times$ \flux. 
     } 
     \label{fig02}
  \end{figure}
  \begin {figure}
   \resizebox{\hsize}{!}{\includegraphics{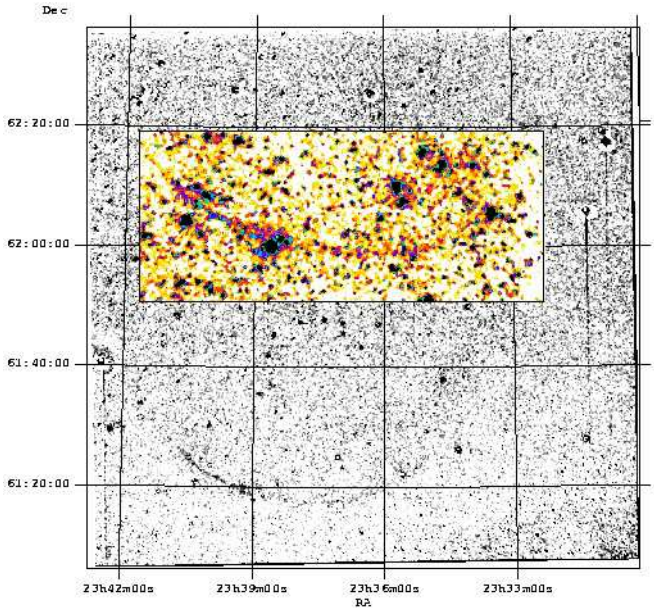}}
    \caption{ \gsnr\ imaged with the medium ionization line of \oiii 5007\AA. 
     The image has been smoothed to suppress the residuals 
     from the imperfect continuum subtraction.
     Shadings run linearly from 0 to 7 $\times$ \flux, while the inlay 
     shows the weak but filamentary nature of the emission present only 
     in the south.
     } 
     \label{fig03}
  \end{figure}
\subsection{The \hnii\ and \sii\ line images}
The low surface brightness of \gsnr\ in the radio, optical and X--ray bands 
seems to be one of its major characteristics. Especially, in the latter band
the emission, if any, is below the ROSAT detection limit. Our \hnii\ 
and \sii\ images (Fig. \ref{fig01} and \ref{fig02},
respectively) reveal weak diffuse emission in the south--east, 
south--west and central areas of the remnant. The bright, extended source LBN 565 
(Lynds \cite{lyn65}) which is believed to be a background source 
(Reich \& Braunsfurth \cite{rei81})  is also visible in 
the low ionization images. In Table~\ref{fluxes}, we list typical  
fluxes measured in several locations within the field of \gsnr\ 
including the \HII\ region LBN 565.  
Since the images are flux calibrated, the \hnii\ and \sii\  frames can be used 
to study the nature of the detected emission. These provide evidence 
that the emission in the south--east and south--west originates from 
shock heated gas (\sii/\ha\ $\sim$0.6-1.0), while LBN 565 is an \HII\ region 
(\sii /\ha\ $\sim$0.3-0.4). Towards the center of the remnant we estimate 
\sii /\ha\ $\sim$0.4-0.6 showing that the diffuse emission may be associated
with \gsnr. We note here that the areas where the 
optical flux is due to shock heating coincide fairly well with the areas 
where the radio emission is strongest, i.e. in the south (see Fig. \ref{fig01}
and Fig. \ref{fig04}).
\subsection{The \oiii\ and \oii\ images}
\begin {figure}
   \resizebox{\hsize}{!}{\includegraphics{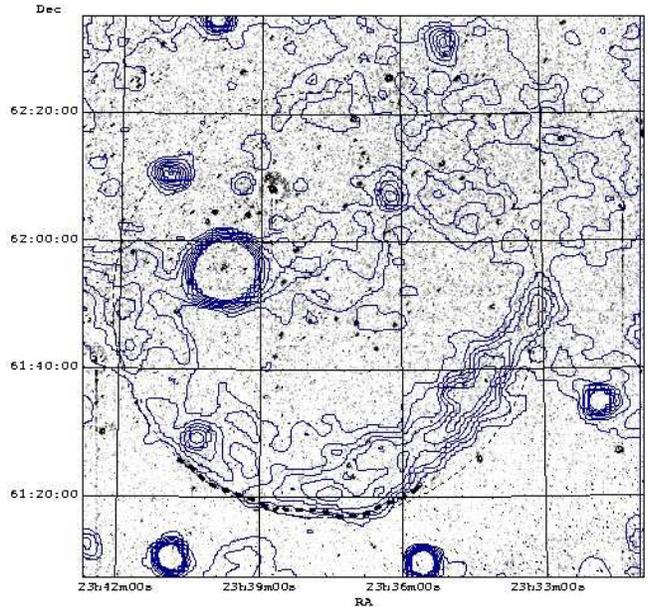}}
    \caption{The correlation between the \oiii\ emission and the radio 
     emission at 4850 MHz is shown in this figure. The thick dashed line 
     outlines the \oiii\ emission, while the thin dashed line defines a 
     circle of 34\arcmin\ radius.  
     The 4850 MHz radio contours (Condon \et\ \cite{con94}) scale linearly 
      from 7$\times$10$^{-4}$ Jy/beam to 0.05 Jy/beam. LBN 565 is also a very
      strong radio source, while it is not detected in \oiii.}
      %%% In 8 steps
     \label{fig04}
  \end{figure}

The morphology of the low ionization line of \oii\ is similar to that 
seen in the \hnii\ and \sii\ images and is not shown here. Typical fluxes 
measured in the \oii\ calibrated image are also given in Table~\ref{fluxes}. 
Contrary to the morphology seen in the low ionization images, the \oiii\ 
morphology is strikingly different. A filament is detected in the south 
which originates at \a\ $\simeq$ 23\h40\m42\s, \dd\ $\simeq$ 
61\degr25\arcmin04\arcsec\ and ends at \a\ $\simeq$ 23\h36\m22\s, \dd\ 
$\simeq$ 61\degr20\arcmin15\arcsec, while some weak emission can be 
seen up to \a\ $\simeq$ 23\h35\m36\s\ and 
\dd\ $\simeq$ 61\degr22\arcmin06\arcsec (Fig. \ref{fig03}). The filament is  not 
continuous over its full extent but there are maxima and minima interchanging \
in the \oiii\ flux, mainly in the south--west. 
This probably reflects inhomogeneities 
in the interstellar ``clouds'' resulting in strong variations of the shock 
velocity upon which the \oiii\ flux depends crucially (Cox \& Raymond 
\cite{cox85}). 
No \oiii\ emission is detected in other areas of the remnant, including 
those areas where diffuse \ha, \nii\ or \sii\ emission is observed. 
The spatial location of the \oiii\ arc like filament nicely matches the 
outermost radio 
contours at 4850 MHz (Fig. \ref{fig04}) suggesting its association with \gsnr. 
The \oiii\ arc probably lies very close to the leading edge of the forward 
shock front. However, the low resolution of the published radio data 
(Condon \et\ \cite{con94}) do not allow us to study in detail this issue. 
\section{The long--slit spectra from \gsnr}
The deep low resolution spectra cover the south--east and south--west 
parts of the remnant. All spectra clearly demonstrate the fact that the 
observed emission must originate from shock heated gas 
(Table~\ref{sfluxes}). We note here that the signal to noise ratios 
quoted in Table~\ref{sfluxes} do not include calibration errors which 
are less than 10\% for the specific days of observation. 
In addition, the ratio of the sulfur lines approaches the low density 
limit indicating low electron densities (e.g. Osterbrock \cite{ost89}). 
The spectra taken from position I (Fig. \ref{fig01}) display strong \oiii\ 
emission relative to \ha\ suggesting shock velocities greater than 
$\sim$100 \vel\ (Cox \& Raymond \cite{cox85}, Hartigan \et\ \cite{har87}), 
while the intensity of the sulfur lines suggests 
electron densities of a few tens of \dens. However, given the errors on the 
individual sulfur fluxes, we estimate that electron densities less than 
$\sim$270 \dens\ are allowed. 
\par
The spectra from position II show roughly the same characteristics with 
those from position I, i.e. shock velocity greater than $\sim$100 \vel\ 
and low electron densities. However, due to the lower absolute fluxes the 
larger errors on the  
sulfur fluxes allow densities up to $\sim$600 \dens. 
The spectra from position III display a different qualitative behavior compared 
to those from positions I and II. Furthermore, we were able to extract 
spectra from different apertures along the slit (IIIa, IIIb, IIIc). 
Spectral variations are observed among these spectra. A significant detection 
of the \oiii\ line is only possible for the spectrum at position IIIa suggesting
shock velocities greater than $\sim$100 \vel. It is found that electron 
densities less than $\sim$400 \dens\ are allowed for this position. 
Emission from the \oiii 5007 \AA\ line is marginally detected in spectrum 
IIIb suggesting shock velocities less than $\sim$100 \vel. Electron 
densities also less than $\sim$400 \dens\ are estimated from the sulfur 
line ratio and its 1$\sigma$ error. Finally, in position IIIc we do not detect 
any \oiii\ emission but for the first time weak \hbeta\ emission is detected. 
  \begin{table*}
        \caption[]{Relative line fluxes}
         \label{sfluxes}
         \begin{flushleft}
         \begin{tabular}{lllllll}
     \hline
 \noalign{\smallskip}
                & I  & II  & IIIa & IIIb & IIIc \cr
\hline
Line (\AA) & F$^{\rm a,b}$ & F$^{\rm a,b}$ & F$^{\rm a,b}$  &  F$^{\rm a,b}$ & 
 F$^{\rm a,b}$ \cr
\hline
4861 \hbeta\   & $<$ 200 & $<$ 240 &  $<$ 140   &  $<$ 180 &  187 (3)$^{\rm c}$   \cr
\hline
4959 [OIII]  & 220 (2)  & --    &  126 (2)  & --  &  --   \cr
\hline
5007 [OIII]  & 688 (5)  & 1508 (5) &   417 (4)& 84 (1) & --  \cr
\hline  
6548 \nii\    & 242 (3)  & --       &   --     & 211 (4) & 190 (4)  \cr
\hline
6563 \ha\     & 1000 (12)& 1000 (4) & 1000 (13)&  1000 (18)& 1000 (17) \cr
\hline
6584 \nii\    & 776 (9)  & 476 (3)  & 690 (8)  &  593 (10) &  602 (12)\cr
\hline
6716 \sii\    & 502 (7)  & 546 (4)  & 725 (11) &  571 (10) &  725 (13)\cr
\hline
6731 \sii\    & 341 (5)  & 371 (3)  & 559 (8)  &  432 (8)  &  530 (12) \cr
\hline
\hline
Absolute \ha\ flux$^{\rm d}$ & 2.4 &   1.5  & 3.0  & 4.2 & 3.3    \cr
\hline
\ha /\hbeta\   & $>$ 5    & $>$ 4.2  & $>$ 7  & $>$ 5.6 & 5.3 ($\pm$ 1.8) \cr
\hline
\sii/\ha\ 	& 0.8  & 0.9 &   1.3  & 1.0   &  1.3 \cr
\hline 
F(6716)/F(6731)	& 1.5 & 1.5 &   1.3  & 1.3  &  1.4\cr
\hline 
\end{tabular}
\end{flushleft}
 ${\rm ^a}$ Uncorrected for interstellar extinction 

${\rm ^b}$ Listed fluxes are a signal to noise weighted
average of the individual fluxes

${\rm ^c}$ Numbers in parentheses represent the signal to noise ratio 
of the quoted fluxes

$^{\rm d}$ In units of \flux\\\
${\rm }$ All fluxes normalized to F(\ha)=1000
\end{table*}
%--------------------------------------------------------
%
\section{Discussion}
The \snr\ \gsnr\ is one of the faintest and least observed remnants in optical
wavelengths (e.g. Boumis \et\ \cite{bou01} on G 17.4-2.3). 
It is also a weak radio source, while extended X-ray emission 
is not detected in the ROSAT band.
The evolved \snr\  under study is now observed for the first time in a 
number of optical emission lines apart from \ha. 
Diffuse emission in the low ionization images is detected in the south and 
central areas of the remnant. Its spatial correlation with regions of strong 
radio emission and the relatively high \sii/\ha\ ratio (\S 3.1) lead us to 
propose that the observed emission is associated with \gsnr. 
We have also discovered a filament in the medium ionization line of \oiii\ in 
the south which is very well correlated with the outermost radio emission at 
4850 MHz. Its projected thickness is $\sim$20\arcsec\ and it is very likely that 
this filament roughly delineates the outer edge of the expanding shell. 
Both the calibrated images and the long--slit spectra 
suggest that it is the result of emission from shock heated gas. 
\gsnr\ belongs to the group of remnants which display strong \oiii\ emission 
in only certain sections of the remnant, e.g. \object{CTB 1} (Fesen \et\ \cite{fes97}), 
\object{CTB~80} (Mavromatakis \et\ \cite{mav01}) and \object{G 17.4-2.3} (Boumis \et\ \cite{bou01}).
The strikingly different morphologies between the low and medium 
ionization lines point to the presence of significant inhomogeneities 
and density variations in the preshock medium. It is known that the 
\ha\ or \sii\ lines are produced in cooler areas behind the shock front, 
while the \oiii\ line is emitted from regions much closer to the front. 
Since in the former case higher column densities are sampled compared to the 
latter, the presence of inhomogeneities and density variations would mainly 
affect the recombination zone where the low ionization lines are produced 
(see Hester \cite{hes87}).  
Such effects may also explain the variations seen in the \oiii/\ha\ ratio 
in the long--slit spectra.  
\par
Emission from the \hbeta\ line is below our detection threshold, while the 
low significance of the \hbeta\ emission detected at position IIIc results in 
an extinction c of 0.75 ($\pm$0.45) or an A$_{\rm V}$ of 1.6 ($\pm$0.9). 
Using the code of Hakkila \et\
(\cite{hak97}) to estimate the interstellar absorption in the direction of 
\gsnr, we find a logarithmic extinction of 1.5 ($\pm$0.3) for a distance 
of 2 kpc and 
2.0 ($\pm$0.3) for a distance of 3 kpc. Our measurement is compatible 
with both estimates within the 2$\sigma$ range due to the large error. 
An extinction in the range of 1.0--1.5 could be more favorable since  
it would imply \hbeta\ fluxes below our detection limit, although other 
values cannot be excluded.
\par
In order to obtain quantitative estimates of basic \snr\ parameters we will 
use the relations 
\begin{equation}
{\rm n_{[\ion{S}{ii}]} \simeq\ 45\ n_c \times V_{s,100}^2},
\end{equation}
and
\begin{equation}
{\rm E_{51}} = 2 \times 10^{-5} \beta^{-1}\ {\rm n_c}\ V_{\rm s,100}^2 \ 
{\rm r_{s,pc}}^3,  
\end{equation}

given by Fesen \& Kirshner (\cite{fes80}) and McKee \& Cowie 
(\cite{mck75}), respectively. The factor $\beta$ is of the order of 
1--2, ${\rm n_{[\ion{S}{ii}]}}$ is the electron density derived from the sulfur line 
ratio, n$_{\rm c}$ is the preshock cloud density, ${\rm E_{51}}$ is the
explosion energy in units of 10$^{51}$ erg, V$_{\rm s,100}$ is the shock velocity
into the clouds in units of 100 \vel, and  {\rm r$_{\rm s,pc}$} the radius of the 
remnant in pc.
The long--slit spectra show that an upper limit on the electron 
density is $\sim$500 \dens\ and we use it in Eq. 1 to obtain 
n$_{\rm c}$ $\times$ V$_{s,100}^2$ $<$ 11. Using this last result in 
Eq. 2, we find  E$_{\rm 51}$ $<$ 1.6 D$_{\rm 2.5 kpc}^3$, a rather coarse 
upper limit. The distance to the remnant in units of 2.5 kpc is denoted 
by D$_{\rm 2.5 kpc}$ (Kulkarni \et\ \cite{kul93}).
\par
Since a direct measurement of the interstellar hydrogen density is not
available, a density of 0.1 \dens\ (F\"urst \et\ \cite{fur93}, Frail \et\ 
\cite{fra94}) and an E$_{\rm 51}$ of 1 and 0.1 will be assumed in the following
calculations. According to Cioffi \et\ (\cite{cio88}) and the above range of
parameters the onset of the pressure--driven snow-plow phase (PDS) would occur 
at 38 pc and 19 pc, respectively. These radii when compared with the 
current radius of $\sim$25 pc at 2.5 kpc suggest that in the former case the 
onset of the PDS phase will occur in another $\sim$40000 yrs, 
while the PDS phase has already started in the latter case. 
We note here that an interstellar density of e.g. 0.5 \dens\ would lower the 
limiting radii to 14 pc and 10 pc, respectively, implying that the remnant is 
well within the radiative phase of its evolution. 
Since neither the distance, nor the ambient medium density are accurately
known, we cannot confidently determine the current stage of evolution of 
\gsnr. However, since the majority of the cases we have examined results 
in radii lower than the current radius it may be possible that the remnant 
has passed the adiabatic phase of its evolution and has entered the snow--plow
phase.
For all these cases, the forward shock velocity at the onset of the radiative
snow--plow phase lies in the range of 250 -- 350 \vel. 
\par
It is proposed that now after $\sim$ 41000 yrs (Kulkarni \et\ \cite{kul93}) 
following the supernova explosion which gave birth to {PSR~2334+61}, 
the remnant is probably in the radiative phase of its evolution.  
Its current radius is 25 D$_{\rm 2.5 kpc}$ pc and the shock front is 
propagating in a inhomogeneous interstellar medium as the optical data 
indicate. Soft X--ray emission from \gsnr\ is not detected in the  
ROSAT observation of PSR 2334+61 (Becker \et\ \cite{bec96}) and we obtain 
an upper limit on the unabsorbed soft X--ray luminosity of 
$\sim$6 $\times$ 10$^{\rm 35}$ erg s$^{-1}$ (0.1 -- 2.0 keV). 
Deep X--ray observations would help to actually determine the X--ray 
properties of the remnant or place new, more stringent upper limits.
\section{Conclusions}
The supernova remnant \gsnr\ was observed in major optical lines. 
New diffuse structures were detected in the south--east and central 
areas of the remnant. The flux calibrated images imply that the 
diffuse emission is associated with \gsnr. 
In addition, we discovered an \oiii\ filament 
in the south which is very well correlated with the radio contours 
at 4850 MHz. Long--slit spectra along this filament suggest that the 
emission arises from shock heated gas, while the sulfur line ratios 
indicate low electron densities.
\begin{acknowledgements}
\end{acknowledgements}
The authors would like to thank R. Fesen for his helpful comments.
We would also like to thank the referee for his remarks and 
suggestions.  
Skinakas Observatory is a collaborative project of the University of
Crete, the Foundation for Research and Technology-Hellas and
the Max-Planck-Institut f\"ur Extraterrestrische Physik.
This work has been supported by a P.EN.E.D. program of the General 
Secretariat of Research and Technology of Greece. 
This research has made use of data obtained through the High Energy 
Astrophysics Science Archive Research Center Online Service, 
provided by the NASA/Goddard Space Flight Center.
%

%----------------

\begin{thebibliography}{}
	\bibitem[1996]{bec96} Becker W., Brazier K.T.S., Tr\"umper J. 1996,
		A\&A 306, 464
		 
	\bibitem[2001]{bou01} Boumis P., Mavromatakis F., Paleologou E.V. 2001, 
		A\&A, submitted

	\bibitem[1988]{cio88} Cioffi D. F., McKee C. F., Bertschinger E.
		1988, ApJ 334, 252

	\bibitem[1994]{con94} Condon J. J., Broderick J. J., Seielstad G. A., 
		Douglas K., Gregory P. C. 1994, 
		AJ 107, 1829

	\bibitem[1985]{cox85} Cox D. P., Raymond J. C. 1985,
		ApJ 298, 651

	\bibitem[1972]{cox72} Cox D. P. 1972, ApJ 178, 159
	
        \bibitem[1980]{fes80} Fesen R. A., Kirshner R. P. 1980
     		ApJ 242, 1023

	\bibitem[1997]{fes97} Fesen R. A., Winkler P.F., Rathore Y., 
		Downes R.A., Wallace D., Tweedy R.W. 1997, 
		AJ 113, 767  

	\bibitem[1994]{fra94} Frail D. A., Goss W. M., Whiteoak J. B. Z. 
		1994, ApJ 437, 781 
		
	\bibitem[1993]{fur93} F\"urst E., Reich W., Seiradakis J. 1993, 
		A\&A 276, 470

	\bibitem[1987]{har87} Hartigan P., Raymond J. and 
		Hartmann L. 1987, ApJ 316, 323
	
	\bibitem[1997]{hak97} Hakkila J., Myers J. M., Stidham B. J. and
		Hartmann D. H., 1997, AJ 114, 2043
	 
	\bibitem[1987]{hes87} Hester J. 1987, ApJ 314, 187
	
	\bibitem[1980]{kal80} Kallas E. and Reich W. 1980, A\&AS 42, 227
	 
	\bibitem[1993]{kul93} Kulkarni S.R., Predehl P., Hasinger G., 
	 	Aschenbach B. 1993, Nature 362, 135 
	 
        \bibitem[1965]{lyn65} Lynds B. T. 1965, ApJS 12, 163  

	\bibitem[2000]{mav00} Mavromatakis F., Papamastorakis J., Paleologou 
		E. V., and Ventura J. 2000, A\&A 353, 371
		
	\bibitem[2001]{mav01} Mavromatakis F., Ventura J., Paleologou E. V., 
		and Papamastorakis J. 2001, A\&A 371, 300
		
	\bibitem[1975]{mck75} McKee C. F., Cowie L. 1975,
		ApJ 195, 715

        \bibitem[1989]{ost89} Osterbrock D. E. 1989,  Astrophysics of 
	           gaseous nebulae, W. H. Freeman \& Company  
     

	\bibitem[1981]{rei81} Reich W. and Braunsfurth E. 1981, A\&A 99, 17
%
\end{thebibliography}
\end{document}